\documentclass[12pt,epsfig]{article}  
\usepackage{amsmath}
\usepackage{amssymb}
\usepackage{graphicx}
\usepackage[left=3.0cm,right=2.0cm, top=3.0cm,bottom=3.0cm]{geometry}
\usepackage{cite}

\begin{document}

\newcommand{\sheptitle}
{Parametrization of lepton mixing matrix in terms of deviations from bi-maximal 
and tri-bimaximl mixing}
\newcommand{\shepauthor}
{Chandan Duarah$^{a,}$\footnote{\it{Corresponding author:} chandan.duarah@gmail.com},
                                  K. Sashikanta Singh$^b$ and N. Nimai Singh$^{b, c}$}
\newcommand{\shepaddress}
   { $^{a}$Department of Physics, Dibrugarh University,
               Dibrugarh - 786004, India \\
   $^b$Department of Physics, Gauhati University, Guwahati - 781014, India\\
   $^c$Department of Physics, Manipur University, Canchipur, Imphal - 795003, India }
\newcommand{\shepabstract}
{We parametrize lepton mixing matrix, known as PMNS matrix, in terms of three 
parameters which account deviations of three mixing angles from their bi-maximal or
tri-bimaximal values. On the basis of this parametrization we can determine
corresponding charged lepton mixing matrix in terms of those three
parameters which can deviate bi-maximal or tri-bimaximal mixing. We find that
the charged lepton mixing matrices which can deviate bi-maximal mixing matrix 
and tri-bimaximal mixing matrix exhibit similar structures. Numerical analysis
shows that these charged lepton mixing matrices are close to CKM matrix of quark
sector.\\
 
Key-words: Lepton mixing matrix, charged lepton correction, Bimaximal mixing and Tri-bimaximal mixing.\\
PACS number: 14.60 Pq}
%888888888888888888888888888888888888888888888888888888888888888888888888
\begin{titlepage}
\begin{flushright}
%hep-ph/yymmnnn
\end{flushright}
\begin{center}
{\large{\bf\sheptitle}}
\bigskip\\
\shepauthor
\\
\mbox{}\\
{\it\shepaddress}\\
\vspace{.5in}
{\bf Abstract}
\bigskip
\end{center}
\setcounter{page}{0}
\shepabstract
\end{titlepage}

%%%%%%%%%%%%%%%%%%55555555555555555555555

\section{Introduction}

Over the last three years contributions from reactor \cite{dbay,reno,dchoo}, accelerator 
\cite{minos,t2k} and solar \cite{sk}
neutrino experiments have provided precise values of three mixing angles and two 
mass squared differences under a three-neutrino mixing scenario. Global analysis 
\cite{fogli,valle,valle1}
of $3\nu$ oscillation data available from various experiments provides us an overall
view on mixing parameters.  \\

As neutrino experiments have been trying for more and more precision measurements of 
neutrino mixing parameters, meanwhile theorists have been trying
to realize the flavour mixing pattern of leptons. Bimaximal mixing (BM) \cite{bm} and 
Tri-bimaximal mixing (TBM) \cite{tbm} have been playing an attractive role in the search 
of flavour mixing pattern over a decade. Both these mixing schemes are
$\mu-\tau$ symmetric \cite{mtau} and predict
maximal atmospheric mixing and zero reactor angle. They differ in their
predictions of solar angle in such that BM mixing predicts maximal value of solar angle
while TBM mixing leads to a value which equals $\arcsin(\frac{1}{\sqrt{3}})$.
Out of these two mixing schemes predictions of TBM mixing are more closer
to global data \cite{fogli,valle,valle1} compared to the other.  
With the confirmation of non zero $\theta_{13}$ the deviation of lepton mixing from
exact BM or TBM pattern is clear. It is therefore useful to study the deviations
of lepton mixing from exact BM or TBM pattern. Deviations from BM or TBM mixing is in fact
a natural idea frequently discussed in the literature 
\cite{bmdev,bmdevgut,tbmdev}. \\ 

In this paper we introduce three
parameters which account for deviations of the three mixing angles, namely solar,
atmospheric and reactor angle from their exact BM or TBM values. We then parametrize
the lepton mixing matrix in terms of these three deviation parameters.
Parametrization of lepton mixing matrix in terms of deviation parameters
is also discussed in Ref. \cite{king}. Our parametrization set up is however
different from that. We mainly implicate the parametrization set up in predicting
possible structure of charged lepton mixing matrix which in turn can generate
the lepton mixing matrix from BM or TBM neutrino mixing via charged lepton
correction. Charged lepton correction \cite{clep,clep1}
 is a very common tool to deviate
special mixing schemes like BM or TBM mixing. Corrections to special mixing
schemes can also be accounted in mass matrix formalism.
We also analyse numerically the charged lepton mixing matrices with an interest
to compare them with the CKM matrix \cite{ckm} of quark sector. In Grand Unified Theory (GUT) 
based models \cite{gut} CKM like charged lepton corrections to special mixing 
schemes are naturally considered. Such models also incorporates 
Quark-Lepton Complementarity (QLC) \cite{qlc}.\\

Rest of the paper is organized as follows : in Section 2 we discuss the
parametrization of the lepton mixing matrix in terms of deviation parameters.
In Section 3 we discuss an implication of our model in charged lepton correction
scenario. Finally Section 4 is devoted to summary and discussion. 

\section{Parametrization of lepton mixing matrix}

In general, lepton mixing matrix, known as PMNS matrix, is parametrized in terms of
three mixing angles, namely $\theta_{12}$, $\theta_{23}$ and $\theta_{13}$ which are
commonly known as solar, atmospheric and reactor angle; and three CP violating 
phases- one Dirac CP phase $\delta$ and two Majorana phases $\alpha$ and $\beta$.
In the standard Particle Data Group (PDG) parametrization it looks like
\begin{equation}
       U_{PMNS} = \begin{pmatrix}
    c_{12} c_{13}                       & s_{12} c_{13} 
                                                          & s_{13} e^{-i \delta}\\
    -s_{12} c_{23}-c_{12} s_{23} s_{13}e^{i \delta} & c_{12} c_{23}-s_{12} s_{23} s_{13} e^{i \delta}
                                                          & s_{23} c_{13}\\
    s_{12} s_{23}-c_{12} c_{23} s_{13}e^{i \delta} & -c_{12} s_{23}-s_{12} c_{23} s_{13} e^{i \delta} 
                                                          & c_{23} c_{13} \\ 
                                                      \end{pmatrix}. P,
\end{equation}
where $c_{ij}=\cos \theta_{ij}$, $s_{ij}=\sin \theta_{ij}$ ($i, j=1,2$) and
$P=diag(1, e^{i \alpha}, e^{i \beta})$ contains the Majorana CP phases.
In the present work we however drop Majorana phase matrix $P$ assuming
that neutrinos obey Dirac nature. \\

Both BM and TBM matrices predict $\theta_{13}^{bm/tb}=0$ and 
$\theta_{23}^{bm/tb}=45^{\circ}$ (suffices $bm$ and $tb$ represent BM and TBM respectively).
However their predictions for solar angle are
different and are given by $\theta_{12}^{bm}=45^{\circ}$ and 
$\theta_{12}^{tb}=\arcsin(\frac{1}{\sqrt{3}})$. Putting these predictions in Eq.(1),
BM and TBM matrices can be obtained as 
\begin{equation}
U_{BM}=\begin{pmatrix}
                   \frac{1}{\sqrt{2}} & \frac{1}{\sqrt{2}} & 0 \\
                    - \frac{1}{2} & \frac{1}{2} & \frac{1}{\sqrt{2}} \\
                     \frac{1}{2} & -\frac{1}{2} & \frac{1}{\sqrt{2}} \\
       \end{pmatrix},
 \end{equation}
 \begin{equation}
        U_{TBM}=\begin{pmatrix}
                  \sqrt{\frac{2}{3}} & \frac{1}{\sqrt{3}} & 0 \\
                 - \sqrt{\frac{1}{6}} & \frac{1}{\sqrt{3}} & \frac{1}{\sqrt{2}} \\
                 \sqrt{\frac{1}{6}} & -\frac{1}{\sqrt{3}} & \frac{1}{\sqrt{2}} \\
                                         \end{pmatrix}.
\end{equation}

We now introduce three parameters which account for the deviations of
three mixing angles from their corresponding BM or TBM values as follows :\\
\begin{equation}
\left. \begin{array}{rcl}
\theta_{12} &=& \theta_{12}^{bm/tb}+\delta\theta_{12}^{bm/tb}, \\
\theta_{23} &=& \theta_{23}^{bm/tb}+\delta\theta_{23}^{bm/tb},\\
\theta_{13} &=& \theta_{13}^{bm/tb}+\delta\theta_{13}^{bm/tb}, \\
\end{array} \right\}
\end{equation}
where the deviation parameters $\delta\theta_{12}^{bm/tb}$ and $\delta\theta_{23}^{bm/tb}$
can take positive as well as negative values whereas $\delta\theta_{13}^{bm/tb}$
takes only positive values. We present the best fit and $3\sigma$ values of mixing angles
and Dirac CP phase in Table 1 \cite{valle1}. Based on these global data we calculate the 
values of deviation parameters and are presented in Table 2.\\

\begin{table}[h]
\centering
 \begin{tabular}{cllll}
\hline
       Model & Parameter & Best fit & 3 $\sigma$  \\
%           \cline{3-4}
\hline
           & $\theta_{12}$ & $34.6^{\circ}$ & $31.8^{\circ}$ - $37.8^{\circ}$ \\
         NH & $\theta_{23}$ & $48.9^{\circ}$ & $38.8^{\circ}$ - $53.3^{\circ}$    \\
         & $\theta_{13}$ & $8.6^{\circ}$ & $7.9^{\circ}$ - $9.3^{\circ}$  \\
         & $\delta$ & $254^{\circ}$ & $0^{\circ}$-$360^{\circ}$ \\
\hline              
    & $\theta_{12}$ & $34.6^{\circ}$ & $31.8^{\circ}$ - $37.8^{\circ}$ \\
  IH & $\theta_{23}$ &  $49.2^{\circ}$ & $39.4^{\circ}$ - $53.1^{\circ}$    \\
      & $\theta_{13}$ & $8.7^{\circ}$ & $8.0^{\circ}$ - $9.4^{\circ}$  \\
       & $\delta$ & $266^{\circ}$ & $0^{\circ}$ - $360^{\circ}$ \\
\hline 
\end{tabular}
\caption{Best fit and $3\sigma$ values of mixing angles and Dirac CP phase
 for normal and inverted hierarchy (NH and IH) from global data \cite{valle1}.}    
\end{table}

For BM mixing we have from Eq.(4)
\begin{equation}
\left. \begin{array}{rcl}
\theta_{12} &=& 45^{\circ}+\delta\theta_{12}^{bm}, \\
\theta_{23} &=& 45^{\circ}+\delta\theta_{23}^{bm}, \\                                                                                                                                                                                                                                                                                                                                                                                                                                                                                                                                                                                                                                                                                                                                                                                                                                                                                                                                                                                
\theta_{13} &=& \delta\theta_{13}^{bm}. \\
\end{array} \right\}
\end{equation}
Substituting these values in Eq.(1) we have PMNS matrix as 
\begin{equation}
U_{PMNS} =  \begin{pmatrix}
	           \frac{1}{\sqrt{2}}p \tilde{r}
	                   & \frac{1}{\sqrt{2}}\tilde{p} \tilde{r}
	                              &  r e^{-i\delta} \\
	  -\frac{1}{2}\left(\tilde{p} \tilde{q} + p q r e^{i\delta}\right) 
	      & \frac{1}{2}\left(p \tilde{q} - \tilde{p} q r e^{i\delta}\right)
	               & \frac{1}{\sqrt{2}} q \tilde{r}  \\
	     \frac{1}{2}\left(\tilde{p} q - p \tilde{q} r e^{i\delta}\right) 
	    & -\frac{1}{2}\left(p q + \tilde{p} \tilde{q} r e^{i\delta}\right) 
	                 & \frac{1}{\sqrt{2}} \tilde{q} \tilde{r}\\
	\end{pmatrix},
\end{equation}
		where 	
\begin{equation}
\left. \begin{array}{rcl}
p &=& \cos \delta\theta_{12}^{bm} - \sin \delta\theta_{12}^{bm} , \\
\tilde{p} &=& \cos \delta\theta_{12}^{bm} + \sin \delta\theta_{12}^{bm} , \\
q &=& \cos \delta\theta_{23}^{bm} + \sin \delta\theta_{23}^{bm}, \\ 
\tilde{q} &=& \cos \delta\theta_{23}^{bm} - \sin \delta\theta_{23}^{bm}, \\
r &=& \sin \delta\theta_{13}^{bm}, \\ 
\tilde{r} &=& \cos \delta\theta_{13}^{bm}. \\
\end{array} \right\}
\end{equation}

For TBM mixing we have from Eq.(4)
\begin{equation}
\left. \begin{array}{rcl}
\theta_{12} &=&  35.26^{\circ}+\delta\theta_{12}^{tb}, \\ 
\theta_{23} &=& 45^{\circ}+\delta\theta_{23}^{tb}, \\                                                                                                                                                                                                                                                                                                                                                                                                                                                                                                                                                                                                                                                                                                                                                                                                                                                                                                                                                                                
\theta_{13} &=& \delta\theta_{13}^{tb}. \\
\end{array} \right\}
\end{equation}
Substituting these values in Eq.(1) we have PMNS matrix as 
\begin{equation}
U_{PMNS} =  \begin{pmatrix}
	           \frac{\sqrt{2}}{\sqrt{3}}p^{\prime} \tilde{r}^{\prime} 
	                   & \frac{1}{\sqrt{3}}\tilde{p}^{\prime} \tilde{r}^{\prime}
	                              &  r^{\prime} e^{-i\delta} \\
	  -\frac{1}{\sqrt{6}}\left(\tilde{p}^{\prime} \tilde{q}^{\prime} 
	  + \sqrt{2}p^{\prime} q^{\prime} r^{\prime} e^{i\delta}\right) 
	      & \frac{1}{\sqrt{3}}\left(p^{\prime} \tilde{q}^{\prime} 
	       - \frac{1}{\sqrt{2}}\tilde{p}^{\prime} q^{\prime} r^{\prime} e^{i\delta}\right)
	               & \frac{1}{\sqrt{2}} q^{\prime} \tilde{r}^{\prime}  \\
	     \frac{1}{\sqrt{6}}\left(\tilde{p}^{\prime} q^{\prime}
	     -\sqrt{2} p^{\prime} \tilde{q}^{\prime} r^{\prime} e^{i\delta}\right) 
	    & -\frac{1}{\sqrt{3}}\left(p^{\prime} q^{\prime}
	              + \frac{1}{\sqrt{2}} \tilde{p}^{\prime} \tilde{q}^{\prime} r^{\prime} e^{i\delta}\right) 
	                 & \frac{1}{\sqrt{2}} \tilde{q}^{\prime} \tilde{r}^{\prime}\\
	\end{pmatrix},
\end{equation}
where 	
\begin{equation}
\left. \begin{array}{rcl}
p^{\prime} &=& \cos \delta\theta_{12}^{tb} -\frac{1}{\sqrt{2}} \sin \delta\theta_{12}^{tb} , \\ 
\tilde{p}^{\prime} &=& \cos \delta\theta_{12}^{tb} + \sqrt{2} \sin \delta\theta_{12}^{tb}, \\
q^{\prime} &=& \cos \delta\theta_{23}^{tb} + \sin \delta\theta_{23}^{tb} , \\ 
\tilde{q}^{\prime} &=& \cos \delta\theta_{23}^{tb} - \sin \delta\theta_{23}^{tb}, \\
r^{\prime} &=& \sin \delta\theta_{13}^{tb} , \\
\tilde{r}^{\prime} &=& \cos \delta\theta_{13}^{tb} . \\
\end{array} \right\}
\end{equation}

\begin{table}
\centering
 \begin{tabular}{cllll}
\hline
      Mixing Scheme &  Model & Parameter & Best fit  & 3 $\sigma$  \\
%           \cline{3-4}
\hline
        &     & $\delta\theta_{12}$ & $-10.4^{\circ}$ & $13.2^{\circ}$ - $(-7.2^{\circ})$ \\
        &  NH & $\delta\theta_{23}$ & $3.9^{\circ}$ & $-6.2^{\circ}$ -  $8.3^{\circ}$    \\
        &     & $\delta\theta_{13}$ & $8.6^{\circ}$ & $7.9^{\circ}$ - $ 9.3^{\circ}$  \\
 BM     & \cline{2-4}               
       &     & $\delta\theta_{12}$ & $-10.4^{\circ}$ & $13.2^{\circ}$ - $( -7.2^{\circ})$ \\
      &  IH & $\delta\theta_{23}$ &  $4.2^{\circ}$ & $-5.6^{\circ}$ - $ 8.1^{\circ}$    \\
       &     & $\delta\theta_{13}$ & $8.7^{\circ}$ & $8.0^{\circ}$ - $ 9.4^{\circ}$  \\
\hline
       &     & $\delta\theta_{12}$ & $-0.66^{\circ}$ & $-3.46^{\circ}$ - $ 2.53^{\circ}$ \\
       &  NH & $\delta\theta_{23}$ & $3.9^{\circ}$ & $-6.2^{\circ}$ - $ 8.3^{\circ}$    \\
       &      & $\delta\theta_{13}$ & $8.6^{\circ}$ & $7.9^{\circ}$ - $ 9.3^{\circ}$  \\
 TBM &  \cline{2-4}               
	  &    & $\delta\theta_{12}$ & $-0.66^{\circ}$ & $-3.46^{\circ}$ - $ 2.53^{\circ}$ \\
      &  IH & $\delta\theta_{23}$ & $4.2^{\circ}$ & $-5.6^{\circ}$ - $ 8.1^{\circ}$    \\
      &  & $\delta\theta_{13}$ & $8.7^{\circ}$ & $8.0^{\circ}$ - $ 9.4^{\circ}$  \\

\hline 
\end{tabular}
\caption{Calculated values of deviation parameters from global data.}    
\end{table}

We want to emphasize that parametrization of lepton mixing matrix in terms of deviation 
parameters has also been discussed by King \cite{king}. There also
exists some interest in parametrizing the lepton mixing matrix in terms of
Wolfenstein parameter $\lambda$ \cite{lamda}, where $\lambda$ accounts for the deviations
of mixing angles from their values predicted by special mixing schemes.

\section{An implication of the model : charged lepton mixing matrix}

Deviations from BM or TBM mixing can be accounted in terms of charged lepton
corrections \cite{clep,clep1}. In the basis where both charged 
lepton mass matrix ($m_l$) and left handed
Majorana mass matrix ($m_{\nu}$) are non diagonal, lepton mixing matrix is given by the product
of two mixing matrices as
\begin{equation}
U_{PMNS}= U_{lL}^{\dagger} U_{\nu},
\end{equation} 
where $U_{lL}$ diagonalizes $m_l$ and $U_{\nu}$ corresponds to the 
diagonalization of $m_{\nu}$. In the basis in which charged lepton mass matrix 
is itself diagonal
PMNS matrix is directly given by $U_{\nu}$, $U_{lL}$ being identity matrix.
The general idea of charged lepton correction is to work in the basis 
where both $m_l$ and $m_{\nu}$ are non diagonal and then considering
$U_{\nu}$ be a special mixing matrix like BM or TBM a small perturbation
to it is accounted from  $U_{lL}$ leading to the desired PMNS matrix. 
Following this set up charged lepton corrections to
special mixing patterns like BM, TBM, Hexagonal mixing etc. are done.
For example charged lepton corrections to BM mixing are found in Refs. \cite{clepbm,clepbmtbm} 
and those to TBM mixing are discussed in Refs. \cite{clepbmtbm,cleptbmgut}. 
With the same idea, in our work, we first find out $U_{lL}$ which can deviate 
BM neutrino mixing matrix and yield
the lepton mixing matrix in Eq.(6). In that case $U_{\nu}$ in Eq.(11) is given by
$U_{BM}$ and corresponding $U_{lL}$ is then given by
\begin{equation}
U_{lL}^{bm} = \begin{pmatrix}
                 a & -\frac{1}{\sqrt{2}}(b + z_1) & \frac{1}{\sqrt{2}} (c-z_2) \\
                 \frac{1}{\sqrt{2}}(d+z_3) & \frac{1}{2}(e+z_4) & \frac{1}{2}(f-z_5) \\
                 -\frac{1}{\sqrt{2}}(d-z_3) & \frac{1}{2}(e-z_4) & \frac{1}{2}(f+z_5) \\
                   \end{pmatrix},
\end{equation}
where
\begin{equation}
\left. \begin{array}{rcl} 
a &=& \cos \delta\theta_{12}^{bm} \tilde{r}, \\
b &=& \sin \delta\theta_{12}^{bm} \tilde{q}, \\
c &=& \sin \delta\theta_{12}^{bm} q, \\
d &=& \sin \delta\theta_{12}^{bm} \tilde{r}, \\
e &=& q \tilde{r}, \\
f &=& \tilde{q} \tilde{r}, \\
z_1 &=& \cos \delta\theta_{12}^{bm} q r e^{-i \delta}, \\
z_2 &=& \cos \delta\theta_{12}^{bm} \tilde{q} r e^{-i \delta}, \\
z_3 &=& r e^{i \delta}, \\
z_4 &=& \cos \delta\theta_{12}^{bm} \tilde{q} - \sin \delta\theta_{12}^{bm} q r e^{-i \delta}, \\
z_5 &=& \cos \delta\theta_{12}^{bm} q - \sin \delta\theta_{12}^{bm} \tilde{q} r e^{-i \delta}. \\
\end{array}\right\}
\end{equation}
The parameters $a$-$f$ and $z_1$-$z_5$ are used to express 
the matrix in Eq.(12) in convenient way. \\

For TBM mixing case $U_{\nu}$ in Eq.(11) is given by $U_{TBM}$ and corresponding
$U_{lL}$ is then given by
\begin{equation}
U_{lL}^{tb} = \begin{pmatrix}
                 a^{\prime} & -\frac{1}{\sqrt{2}}(b^{\prime} + z_1^{\prime}) 
                               & \frac{1}{\sqrt{2}} (c^{\prime}-z_2^{\prime}) \\
                 \frac{1}{\sqrt{2}}(d^{\prime}+z_3^{\prime}) 
                 & \frac{1}{2}(e^{\prime}+z_4^{\prime}) & \frac{1}{2}(f^{\prime}-z_5^{\prime}) \\
                 -\frac{1}{\sqrt{2}}(d^{\prime}-z_3^{\prime}) 
                 & \frac{1}{2}(e^{\prime}-z_4^{\prime}) & \frac{1}{2}(f^{\prime}+z_5^{\prime}) \\
                   \end{pmatrix},
\end{equation}
where the parameters $a^{\prime}$-$f^{\prime}$ and $z_1^{\prime}$-$z_5^{\prime}$ are
given by Eq.(13) with the substitutions of $\delta\theta_{12}^{bm}$, $q$, $\tilde{q}$,
$r$ and $\tilde{r}$ by $\delta\theta_{12}^{tb}$, $q^{\prime}$, $\tilde{q}^{\prime}$,
$r^{\prime}$ and $\tilde{r}^{\prime}$ respectively. \\

We note that both charged lepton mixing matrices $U_{lL}^{bm} $ and $U_{lL}^{tb} $
have similar structure due to $\mu-\tau$ symmetry of BM and TBM mixing matrices.
We estimate the numerical values (in modulus) of the elements of these mixing matrices 
for best fit values of deviation parameters and are presented in Eqs. (15) and (16).
\begin{equation}
U_{lL}^{bm} = \begin{pmatrix}
                0.972512 & 0.183349 & 0.143535 \\
                0.185651 & 0.980189 & 0.062209 \\
                0.140544 & 0.074912 & 0.980319 \\
                 \end{pmatrix}.
\end{equation}
\begin{equation}
U_{lL}^{tb} = \begin{pmatrix}
                0.988657 & 0.114991 & 0.096260 \\
                0.108234 & 0.991394 & 0.072972 \\
                0.103806 & 0.062329 & 0.992184 \\
                 \end{pmatrix}.
\end{equation}

Naturally there exists naive interest in searching connection between quark sector 
and lepton sector. Grand unified theories (GUTs) generally provide the framework 
for quark-lepton unification. Quark-lepton-complementarity (QLC), which signifies interesting 
phenomenological relations between the lepton and quark mixing angles supports
the idea of grand unification. Derivation of QLC relations assumes the deviation 
of lepton mixing from exact BM pattern to be described by quark mixing matrix. 
In GUT based models \cite{bmdevgut,cleptbmgut,gut} charged lepton corrections 
to special neutrino mixing schemes
are considered as CKM like. From such points of view we make comparison of the charged
lepton mixing matrices in Eqs. (15) and (16) with the CKM matrix. For convenience, we present
the best fit values (in modulus) of the elements of CKM matrix in Eq. (17) \cite{ckmdata}.
\begin{equation}
V_{CKM} = \begin{pmatrix}
                0.97428 & 0.2253 & 0.00347 \\
                0.2252 & 0.97345 & 0.0410 \\
                0.00862 & 0.0403 & 0.999152 \\
                 \end{pmatrix}.
\end{equation}

Wee see that both the mixing matrices are close to CKM matrix.
Like CKM matrix the diagonal elements in these mixing matrices are close to unity
and non diagonal elements exhibit an approximate symmetric nature. One significant 
point, we note, is that the corner elements, namely $(U_{lL})_{13}$ and $(U_{lL})_{31}$
in both the mixing matrices are relatively larger compared to those of $V_{CKM}$ matrix.

\section{Summary and discussion}

BM and TBM are two special neutrino mixing schemes. To accommodate non zero $\theta_{13}$
and deviations of solar mixing and atmospheric mixing from maximality these special
mixing schemes should be modified. We have three parameters, viz. $\delta\theta_{12}^{bm/tb}$,
$\delta\theta_{23}^{bm/tb}$ and $\delta\theta_{13}^{bm/tb}$, which account the deviations
of lepton mixing angles from their BM or TBM values. Numerical values of these deviation
parameters can be obtained from global $3\nu$ oscillation data. We then parametrize 
PMNS matrix in terms of these parameters. Such parametrization of lepton mixing
matrix may help authors in phenomenological works which incorporate deviation of
special mixing schemes. We implicate our parametrization set up in predicting 
possible structure of charged lepton mixing matrices which can generate the
desired lepton mixing matrix from BM or TBM mixing matrices. We have found that
charged lepton mixing matrices $U_{lL}$'s in both cases (BM and TBM) exhibit
similar structures. Numerical analysis shows that these mixing matrices 
($U_{lL}^{bm}$ and $U_{lL}^{tbm}$), necessary to deviate BM mixing and TBM 
mixing in obtaining mixing parameters consistent with global data, are close
to the CKM matrix of quark sector. This result is in agreement with the assumption,
generally made in GUT based model, that charged lepton correction
to neutrino mixing can be considered as CKM like.

\end{document}